\documentclass[letterpaper,english,aps,prl,superscriptaddress,twocolumn, showpacs, amsfonts, amssymb]{revtex4}

\usepackage{amsmath}
\usepackage{graphicx}
\usepackage{amssymb}
\usepackage{dsfont}
\usepackage{color}
\usepackage[bookmarks=true,colorlinks,linkcolor=blue,urlcolor=blue,citecolor=blue]{hyperref}

\newcommand{\be}{\begin{equation}}
\newcommand{\ee}{\end{equation}}
\newcommand{\bea}{\begin{eqnarray}}
\newcommand{\eea}{\end{eqnarray}}
\newcommand{\mc}{\mathcal}

\usepackage{hyperref}
\usepackage{tikz}
\usetikzlibrary{calc}

\usepackage{epsfig,graphicx,psfrag,amsmath,amssymb,float}
\usepackage[T1]{fontenc}
\usepackage[latin9]{inputenc}
\usepackage{amsmath}
\usepackage{amssymb}
\usepackage{color}
\usepackage{amscd}
\usepackage{bm}
\usepackage{bbm} 
\usepackage{psfrag}

\begin{document}

\title{Tunable  breakdown of the polaron picture  for  mobile impurities    in a  topological semimetal }

\author{M. A. Caracanhas}
\affiliation{Instituto de  F\'{i}sica de S\~ao Carlos, Universidade de S\~ao Paulo, CP 369, S\~ao Carlos, SP, 13560-970, Brazil}
\affiliation{Institute for Theoretical Physics, Center for Extreme Matter and Emergent Phenomena, Utrecht University, Princetonplein 5, 3584 CC Utrecht, The Netherlands}
\author{R. G. Pereira}
\affiliation{Instituto de F\'{i}sica de S\~ao Carlos, Universidade de S\~ao Paulo, CP 369, S\~ao Carlos, SP, 13560-970, Brazil}
\affiliation{
International Institute of Physics  and 
Departamento de F\'isica Te\'orica e Experimental, Universidade Federal do Rio Grande do Norte, 59078-970 Natal-RN, Brazil}

\date{\today} 
\begin{abstract}
Mobile impurities in cold atomic gases constitute  a  new platform  for investigating  polaron physics. Here we show that when impurity atoms interact with  a two-dimensional Fermi gas with quadratic band touching the polaron picture may either hold or break down depending on the particle-hole asymmetry of the  band structure.  If the hole band has a smaller effective mass than the particle band, the quasiparticle is stable and its  diffusion coefficient  varies with  temperature as $D(T) \propto \ln^2 T$. If the hole band has larger mass, the quasiparticle weight vanishes at low energies due to an emergent orthogonality catastrophe. In this case we map the  problem onto a set of one-dimensional channels and use conformal field theory techniques to obtain $D(T)\propto T^{\nu}$ with an interaction-dependent exponent $\nu$. The different regimes can be detected in the nonequilibrium expansion dynamics of an initially confined impurity. 

\end{abstract}

\pacs{67.85.Lm, 67.85.Pq, 71.38.Fp} 
\maketitle

{\it Introduction.}---Recent experiments with   mixtures of ultracold atoms have rekindled the interest in mobile impurities in quantum many-body systems \cite{SchirotzekPRL2009,PalzerPRL2009,KohstallNature2012,KoschorreckNature2012,SpethmannPRL2012,FukuharaNP2013,HuPRL2016,JorgensenPRL2016}. In these experiments, the mobile impurity is represented by an atom of a dilute  species that interacts with collective excitations of a  majority species, which can be either bosonic \cite{CucchiettiPRL2006,BrudererPRA2007,TemperePRB2009,CaracanhasPRL2013,GrusdtSR2015,ChristensenPRL2015} or fermionic \cite{ChevyPRA2006,CombescotPRL2007,MathyPRL2011,SchmidtPRA2012,MassignanRPP2014}.  In the latter case, the quasiparticle in the interacting system is called a Fermi polaron and is   formed by an atom dressed by density fluctuations of the Fermi gas. The study  of mobile impurities may lead to new techniques to  probe  strongly correlated states of matter \cite{PunkPRA2013,GrusdtarXiv2015}. In addition,  it   allows us to reassess some fundamental questions about the formation of quasiparticles, while performing quantitative tests of existing theories \cite{MassignanRPP2014,ChristensenPRL2015}.


On the other hand, it is natural to ask whether mobile impurities in cold atomic gases  can also be used to investigate  the \emph{breakdown} of the quasiparticle picture. In fact, an outstanding problem in condensed matter physics pertains to the properties of phases without well-defined quasiparticles     \cite{SenthilPRB2008,WitczakKrempaNJP2014,HartnollNP2015}, a famous example  of which  is the strange metal phase of hole-doped  cuprates \cite{LeeRMP2006}. Experience in this field has taught us that one route to the absence of quasiparticles (in the sense of  vanishing quasiparticle weight \cite{VarmaPhysRep2007}) is the coupling of a system to soft fluctuations near a quantum critical point \cite{VarmaPRL1989,AltshulerPRB1994,PolchisnkiNPB1994,NayakNPB1994}.

  In this work, we propose and analyze a mobile impurity model that can be  driven between two regimes, in which the polaron picture either holds or breaks down, by varying a single parameter of a microscopic Hamiltonian. 
 The main idea is to find a scale-invariant model where a marginal interaction  gives rise to infrared singularities analogous to those in theories of non-Fermi liquids \cite{VarmaPhysRep2007}. For this purpose, we must couple the impurity with \emph{quadratic} dispersion to an environment  that also features  quadratic dispersion at low energies. Indeed, the  case of  short-range interactions with  linearly dispersing  critical  modes only involves strictly  irrelevant perturbations  of the free model \cite{PunkPRA2013}. For this reason, we consider  a two-dimensional (2D) Fermi gas with a quadratic band crossing point (QBCP) \cite{SunPRL2009,UebelackerPRB2011,FuPRL2011,SunNP2012,TkachovPRB2013,WangPRB2015}. This peculiar band touching    can   be protected by point group  symmetries when it is associated with a nontrivial Berry flux, as in the checkerboard lattice  \cite{SunPRL2009}, thus characterizing a type of topological semimetal \cite{SunNP2012}. 
For free fermions, tuning the chemical potential  to the QBCP leads to a   low-energy spectrum of  particle-hole pairs with quadratic dispersion. Remarkably,   a repulsive interaction between  fermions in the bulk is a marginally relevant perturbation that drives  instabilities towards nematic or quantum Hall phases \cite{SunPRL2009,UebelackerPRB2011}. 

Our mobile impurity model can be viewed as  the limit of extreme population imbalance of the model in Ref. \cite{SunPRL2009}, in which a single spin-down fermion interacts with a finite density of spin-up fermions  tuned to the 
QBCP. If the interaction is restricted to the $s$-wave channel, there is no direct interaction among spin-up fermions. 
In the following we show that the fate of the mobile impurity 
depends on the particle-hole asymmetry of the bulk fermion bands. In the regime where the filled  band below the QBCP has a smaller effective mass than the empty band above it,  the polaron is well defined, but  the ratio between the decay rate and the  energy vanishes only logarithmically in the low-energy limit. In the opposite regime,  we find  a divergent  enhancement of the impurity mass and vanishing quasiparticle weight  due to an emergent  orthogonality catastrophe (OC) \cite{AndersonPRL1967,KnapPRX2012,mahan2000many,gogolin2004bosonization}. It is possible to switch  between the two regimes by controlling hopping parameters in the optical lattice.  Finally, we show that one can clearly distinguish between these regimes   by measuring the diffusion coefficient in the  expansion dynamics of an initially confined impurity. We stress that the exotic behavior discussed does not occur for an impurity  immersed in a conventional 2D Fermi gas where the low-energy particle-hole pairs have linear dispersion about the Fermi surface \cite{MassignanRPP2014}.

\begin{figure}
  \begin{center}
\includegraphics*[width=.95\columnwidth]{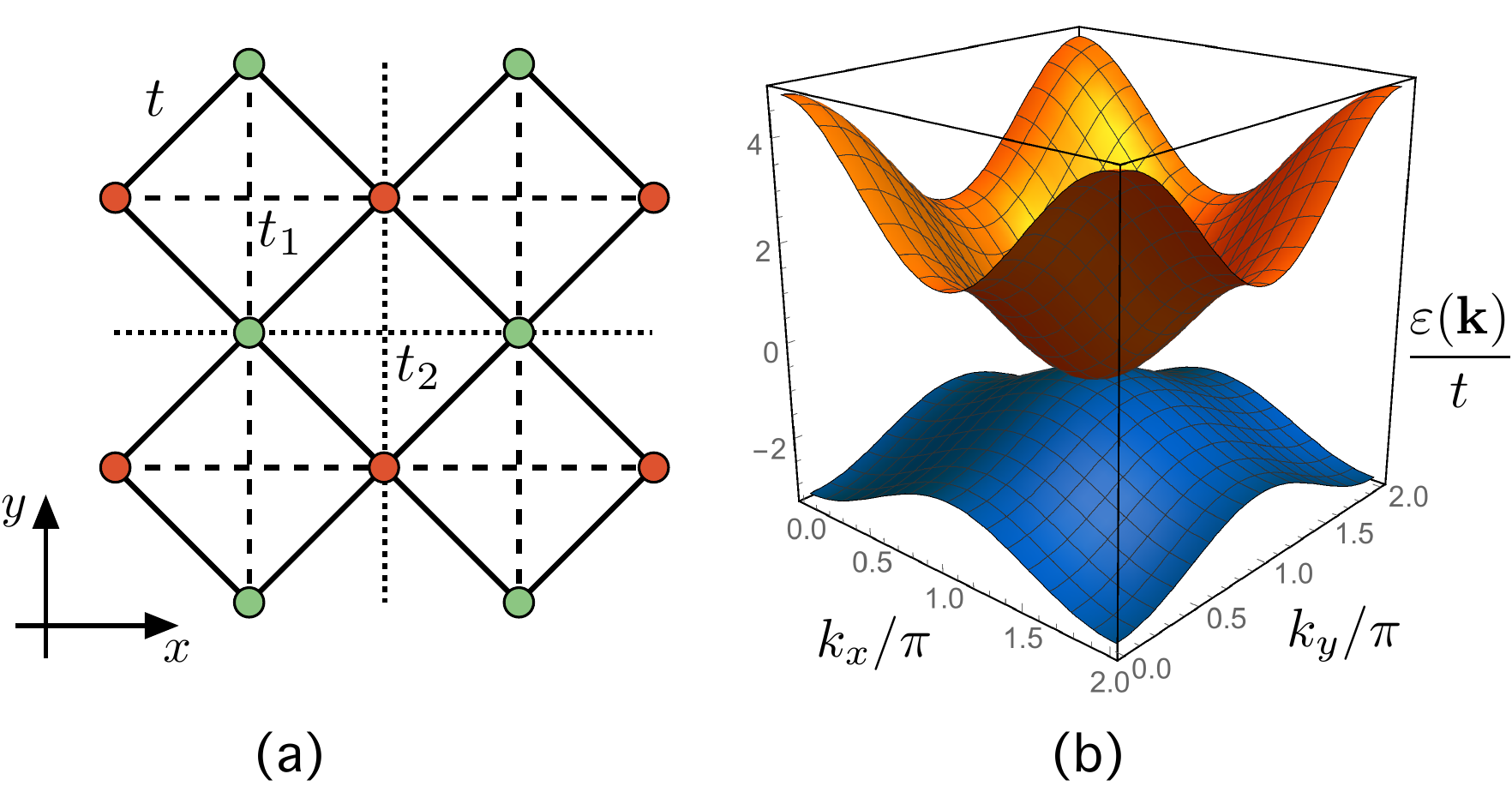}
\end{center}
\caption{(Color online) (a) Checkerboard lattice. Red sites belong to the A sublattice and green sites to the B sublattice. There is a hopping parameter $t$ (solid line) between nearest neighbors in different sublattices and  direction-dependent hopping $t_1$ (dashed line) or $t_2$ (dotted line) between second neighbors.  (b) Band structure  for  $t_1/t=0.7$ and $t_2/t=-0.3$,  showing a QBCP in the  regime $m_+<m_-$.
}\label{fig:lattice}
\end{figure}

{\it Model.}---We consider the lattice model\be
H=\sum_{i,j}(t_{ij}c^\dagger_ic^{\phantom\dagger}_j+J_{ij}d^\dagger_id^{\phantom\dagger}_j)+U\sum_{i}c^\dagger_ic^{\phantom\dagger}_id^\dagger_id^{\phantom\dagger}_i, 
\ee
where  $c_j$ annihilates a fermion of the majority   species  on site $j$ and $d_j$ annihilates the mobile impurity. The Hilbert space obeys the constraint $\sum_jd^\dagger_jd^{\phantom\dagger}_j=1$. The hopping parameters $t_{ij}$ and $J_{ij}$ are defined on a checkerboard lattice with sublattices A and B. For the $c$ fermions, we assign   $t_{ij}$ as illustrated in Fig. \ref{fig:lattice}(a)  \cite{fradkin1991field}.  For the impurity, we set $J_{ij}=rt_{ij}$, where $r$ is the parameter that controls the mass ratio, but the lattice is not crucial for the impurity.  We can also consider a free-particle dispersion if the impurity  does not couple to  a species-specific lattice potential \cite{LeBlancPRA2007}. We note that optical checkerboard lattices have been realized experimentally \cite{OlschlagerPRL2012,WindpassingerRPP2013}. Finally, $U>0$ is the strength of the on-site repulsion, which  is related to the interspecies $s$-wave scattering length \cite{JakschPRL1998}. 

For $U=0$, the tight-binding Hamiltonian for the majority fermions can be diagonalized in the form $H_0=\sum_{i,j}t_{ij}c^\dagger_ic^{\phantom\dagger}_j=\sum_{\mathbf k} \Psi^\dagger_{\mathbf k}H_0(\mathbf k)\Psi^{\phantom\dagger}_{\mathbf k}$, where $\Psi_{\mathbf k}=(a_{\mathbf k},  b_{\mathbf k})^t$ is a two-component spinor and \bea
H_0(\mathbf k)&=&2t_I(\cos k_x+\cos k_y)\mathbbm 1+2t_z(\cos k_x-\cos k_y)\sigma_z\nonumber\\
&&+8t_x\cos(k_x/2)\cos(k_y/2)\sigma_x,\label{H0k}\eea
with $t_I=(t_1+t_2)/2$, $t_z=(t_1-t_2)/2$,  and $t_x=t/2$.
Here the momentum $\mathbf k$ is   in the Brillouin zone of the square lattice  (with lattice parameter set to 1) and the Pauli matrices $\sigma_l$, $l\in\{x,y,z\}$, act in the internal sublattice space of $\Psi_{\mathbf k}$. For $t_I=0$, the Hamiltonian  has a particle-hole symmetry $\mc C^{-1} H_0(\mathbf k)\mc C=-H_0(\mathbf k)$ with $\mc C=\sigma_y$.

We consider the fermion density   at half-filling, $\langle c_j^\dagger c^{\phantom\dagger}_j\rangle=1/2$. At low energies, we expand the Hamiltonian about the band touching point $\mathbf Q=(\pi,\pi)$  and obtain ${H_0(\mathbf Q+\mathbf p)}\approx {t_I(p_x^2+p_y^2-4)}\mathbbm 1+t_z( p_x^2- p_y^2)\sigma_z+2t_xp_x p_y\sigma_x$.  Hereafter we restrict to  $|t_x|=|t_z|$, in which case the model has  continuous rotational invariance in the continuum limit  \cite{SunPRL2009}.  The dispersion relation about the QBCP (dropping a constant) is given by $
\varepsilon_{\pm}(\mathbf p)=\pm   p^2/(2m_\pm)$,
where $m_+$ ($m_-$) is the effective mass of particles (holes) in  the upper (lower) band [see Fig. \ref{fig:lattice}(b)]. In terms of the  parameters in Eq. (\ref{H0k}), $m_{\pm}=(t_z\pm t_I)^{-1}$, where we assume $t_z>t_I\geq0$. As expected, $m_+=m_-$ in the particle-hole symmetric case $t_I=0$, and the sign of $\Delta m=m_+-m_-$ can be controlled by varying $t_I$.

For the free impurity, we expand the Hamiltonian about the lowest-energy state at $\mathbf k=0$ and obtain the dispersion  $E(\mathbf p)\approx p^2/(2M)$. For $J_{ij}=rt_{ij}$, the free impurity mass is    $M=[r(t_z-t_I)]^{-1}$. 

Turning on a weak interaction $U\ll t_z$, we obtain the mobile impurity model in the continuum limit\bea
H_{\text{mob}}&=&\int d^2r\, \left[\Psi^\dagger(\mathbf r)\mc H_0(\mathbf r)\Psi(\mathbf r)-\frac{1}{2M}d^\dagger(\mathbf r)\nabla^2d^{\phantom\dagger}(\mathbf r)\right.\nonumber\\
&&\left.+2\pi\gamma \Psi^\dagger(\mathbf r)\Psi (\mathbf r)d^\dagger (\mathbf r)d(\mathbf r)\right],\label{lowenergyH}
\eea
where $\Psi(\mathbf r)=(a(\mathbf r),b(\mathbf r))^t$  is the fermion field operator, $\mc H_0(\mathbf r)= -t_I\nabla^2\mathbbm 1-t_z[( \partial_x^2-\partial_y^2)\sigma_z+2\partial_x \partial_y\sigma_x]$, and $\gamma \sim \mc O(U)$  is the  coupling constant with units of inverse mass. 
The effective action corresponding to the Hamiltonian in Eq. (\ref{lowenergyH}) is scale invariant with  dynamical exponent $z=2$ \cite{SunPRL2009,fradkin1991field}.    
A bosonic analogue of this model 
appears in the problem of a mobile impurity immersed in a Bose-Einstein condensate  in a vortex-lattice state \cite{CaracanhasPRL2013,CaracanhasPRA2015}. 
An attempt to apply perturbation theory in $\gamma$ reveals logarithmic singularities reminiscent of the behavior in critical  one-dimensional (1D) systems \cite{McGuireJMP1965,CastellaPRB1993,KantianPRL2014}. This is a first sign of the peculiar effects of the marginal impurity-fermion interaction.

{\it Low-energy fixed points.}---We proceed with  a perturbative renormalization group (RG) analysis of the interaction in Eq. (\ref{lowenergyH}).   Computing the effective vertex and impurity self-energy to   $\mc O(\gamma^2)$, we obtain the set of RG equations
\begin{subequations}
\bea
\frac{d\gamma}{d\ell}&=&\frac{\gamma^2(\mu_--\mu_+)Z}4,\label{RGgamma}\\
\frac{dZ}{d\ell}&=&-\frac{\gamma^2\mu_+\mu_-Z}2\, F_Z\left(\frac{m_+}{M},\frac{m_-}{M}\right),\label{RGZ}\\
\frac{dM}{d\ell}&=&  \frac{\gamma^2(\mu_+\mu_-)^{3/2}}2\, F_M\left(\frac{m_+}{M},\frac{m_-}{M}\right).\label{RGM}
\eea
\end{subequations}
Here $\ell=\ln(\Lambda_0/\Lambda)$, with $\Lambda$ the ultraviolet cutoff and $\Lambda_0 \sim t_z$ the bare cutoff. The parameters $\mu_\pm=\left(\frac{1}{M}+\frac1{m_\pm}\right)^{-1}$ are reduced masses, $Z$ is the quasiparticle weight in the impurity Green's function  $G_{d}(\mathbf p,\omega)\approx Z/[\omega-p^2/(2M)]$, and $F_Z(r_+,r_-)$ and $F_M(r_+,r_-)$, with $r_\pm=m_\pm/M$, are  dimensionless functions of the mass ratios that can be determined numerically (see Supplemental Material \cite{supplem}). 

\begin{figure}
  \begin{center}
\includegraphics*[width=.8\columnwidth]{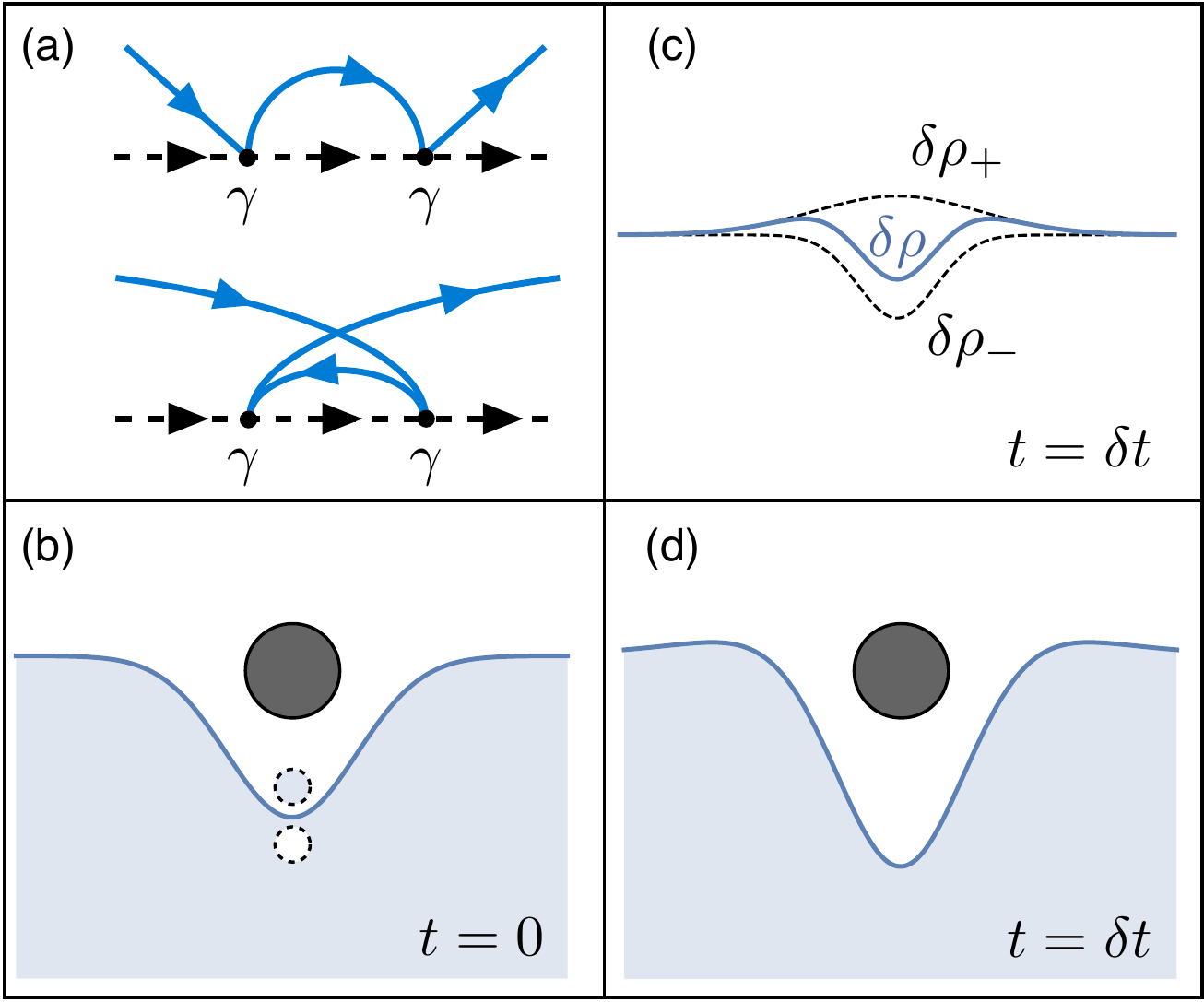}
\end{center}
\caption{(Color online) Renormalization of the impurity-fermion interaction. (a) Diagrams that  contribute  at $\mc O (\gamma^2)$.  The solid line represents fermion propagators and the dashed   line the impurity propagator. Panels (b)-(d) illustrate the   mechanism. (b) For  $\gamma>0$, the fermion density (blue shaded region) is depleted near the impurity. Conversely, the impurity experiences an effective potential that attracts it to the region of lower density. If the impurity is localized, the eigenstates of the Hamiltonian are the scattering states of a static potential. If the impurity mass is large but finite, the impurity  can create particle-hole pairs as it moves slowly through the system. (c)   For $m_+<m_-$, the particle wave packet (with density $\delta \rho_+$) will diffuse faster and have  a larger width than the hole wave packet  (with density $\delta \rho_-$) after some short time $\delta t$.  The net  density $\delta \rho=\delta \rho_++\delta \rho_-$   is then a nonzero function of position with a  minimum at the center. (d) This leads to a further depletion  of the total density at the instantaneous position of the impurity and an enhancement of the effective potential.   For $m_+>m_-$, the sign of $\delta \rho$ is reversed and the effective potential becomes shallower, implying that the interaction  flows to weak coupling.
}\label{fig:RGcartoon}
\end{figure}

Remarkably, Eq. (\ref{RGgamma})  reveals that  the coupling   $\gamma>0$ can be   relevant or irrelevant depending on particle-hole symmetry breaking. For $m_+=m_-$, the particle and hole diagrams that contribute to the renormalization of the interaction vertex at $\mc O(\gamma^2)$, shown in Fig. \ref{fig:RGcartoon}(a), cancel exactly \cite{comentario}. We have verified that this cancellation extends to higher orders in perturbation theory. Thus, in the particle-hole symmetric case the impurity-fermion interaction is strictly marginal. 

For $m_+>m_-$, the interaction becomes marginally irrelevant. In this  regime, the effective coupling   at scale $\Lambda$ vanishes logarithmically, $\gamma(\Lambda) \sim 1/\ln (\Lambda_0/\Lambda)$, for $\Lambda\to 0$. Equations (\ref{RGZ}) and (\ref{RGM}) imply that the  impurity quasiparticle weight $Z(\Lambda)$ initially  decreases and the effective mass $M(\Lambda)$ increases under the RG flow, but they both converge to finite values as $\gamma\to0$. The picture for the low-energy fixed point is a Fermi polaron with renormalized $Z^*$ and $M^*$. In addition, the polaron has a finite decay rate  because the single-particle dispersion is inside a continuum of   particle-hole pairs and there is phase space for scattering at arbitrarily low energies \cite{CaracanhasPRL2013}. On dimensional grounds, the  decay rate for the polaron with momentum $\mathbf p$ is $1/\tau_{\mathbf p}\propto \gamma^2[E(\mathbf p)]E(\mathbf p)\propto p^2/\ln^2 (p)$. Therefore, the quasiparticle is  marginally stable as the ratio between the decay rate and the energy vanishes   as $1/\ln^2(p)$ for $p\to0$. Nonetheless, the impurity spectral function, which can be measured in cold atoms using rf spectroscopy \cite{SchirotzekPRL2009,ShashiPRA89}, must exhibit an approximately Lorentzian quasiparticle peak.

By contrast, the interaction is marginally relevant for $m_+<m_-$.  The picture for the low-energy fixed point in this   regime is the ``self-trapping'' of the  impurity \cite{CaracanhasPRA2015}. As     $M(\Lambda)$ grows without bound, we can analyze the RG equations for $M\gg m_\pm$ using $F_Z(r_+,r_-)\approx 1$ and $F_M(r_+,r_-)\approx \sqrt{\frac{r_+}{r_-}}+ \sqrt{\frac{r_-}{r_+}}$ for $r_\pm\ll1$. The result is that  the effective mass diverges logarithmically, $M(\Lambda)\sim \ln (\Lambda_0/\Lambda)$, while the quasiparticle weight vanishes as a power law, $Z(\Lambda)\sim \Lambda^{g^2/2}$, where $g=\gamma^*(m_+m_-)^{\frac12}$ with $\gamma^*$  the   renormalized coupling in the low-energy limit. Note that  $\gamma(\Lambda)$ does not diverge for $\Lambda\to 0$ because its flow is slowed down by the suppression of $Z(\Lambda)$. This is the behavior expected from the OC for a localized impurity  \cite{AndersonPRL1967,KnapPRX2012,mahan2000many,gogolin2004bosonization}.  We stress that here  the OC occurs in the absence of a Fermi surface. This is possible because the 2D QBCP has a finite density of states.

To demonstrate the OC explicitly, we analyze the vicinity of  the infinite-mass  fixed point.  Neglecting the kinetic energy of the impurity (of order $M^{-1}$), we 
consider the Hamiltonian for the impurity localized at $\mathbf r=0$:\bea
H_{\text{loc}}=\int d^2r \, \Psi^\dagger(\mathbf r)[\mc H_0(\mathbf r)+2\pi \gamma^*\delta(\mathbf r)]\Psi(\mathbf r).\label{Hloc}
\eea
We  use the partial wave expansion\be
\Psi^{\phantom\dagger}_\lambda(\mathbf p)=\Psi^{\phantom\dagger}_\lambda( p,\theta_{\mathbf p})=\frac{1}{\sqrt{2\pi p}}\sum_{\ell \in\mathbb Z} e^{i\ell \theta_{\mathbf p}}\chi_{\lambda\ell}(p),\label{modeexp}
\ee
where $\lambda=\pm$ is the band index, $\theta_{\mathbf p}=\arctan(p_y/p_x)$ is the polar  angle of   $\mathbf p$, and $\chi_{\lambda\ell}(p)$ obeys $\{\chi^{\phantom\dagger}_{\lambda\ell}(p),\chi^\dagger_{\lambda^\prime\ell^\prime}(k)\}=\delta_{\lambda,\lambda^\prime}\delta_{\ell,\ell^\prime}\delta(p-k)$. We then  define the rescaled field $\psi_{\lambda,\ell}(\mc K)=2\pi (m_\lambda/p)^{1/2}\chi_{\lambda,\ell}(p)$, where $\mc K=p^2/(8\pi ^2 m_\lambda)$. This allows us to introduce a  1D chiral fermion for each $\ell$ channel by $
\psi_\ell(x)= \int_{-\infty}^{+\infty} \frac{d\mc K}{\sqrt{2\pi}}\, e^{i\mc K x}[\theta(\mc K)\psi_{+,\ell}(\mc K)+i\text{sgn}(\ell)\theta(-\mc K)\psi_{-,\ell}(-\mc K)]$. 
We can then rewrite  Eq. (\ref{Hloc}) in the form \bea
H_{\text{loc}}&=&\sum_{\ell\in \mathbb Z} \int_{-\infty}^\infty dx\, \psi^\dagger_\ell(-i\partial_x)\psi^{\phantom\dagger}_\ell\nonumber\\
&&+ \pi g[\psi^\dagger_1(0)\psi^{\phantom\dagger}_1(0)+\psi^\dagger_{-1}(0)\psi^{\phantom\dagger}_{-1}(0)].\label{1Dchannels}
\eea
Thus, only the $\ell=\pm1$ channels couple to the impurity potential. This is due to the angular dependence of the eigenstates of $H_0$, which is also responsible for the nonzero Berry flux of the QBCP \cite{supplem}. 

The Hamiltonian for $\ell=\pm1$ channels in Eq. (\ref{1Dchannels}) has the standard form of a 1D conformal field theory with a boundary condition changing operator  \cite{SchottePR1969,AffleckJPA1994}. The boundary term can be eliminated by a unitary transformation that changes the scaling dimension of various operators. We can then compute correlation functions 
by standard methods  \cite{gogolin2004bosonization}. In particular, the impurity Green's function has a power-law decay in time $|G_d(\mathbf p=0,t)|\propto t^{-g^2/2}$. 
In the frequency domain,  the line shape of the spectral function becomes  asymmetric, approaching a power-law singularity as $p\to 0$ \cite{CaracanhasPRA2015}. At finite $p$, the singularity is  broadened  at the energy scale $\sim g^2p^2/M[E(\mathbf p)]$ due to  the recoil of the heavy impurity \cite{GavoretJPhys1969,Doniach1970,RoschAdvPhys1999}. The physical mechanism behind this low-energy fixed point can be understood within  a simple cartoon picture in the limit $M\gg m_{\pm}$ as illustrated in Fig. \ref{fig:RGcartoon}. 

\begin{figure}
  \begin{center}
\includegraphics*[width=.61\columnwidth]{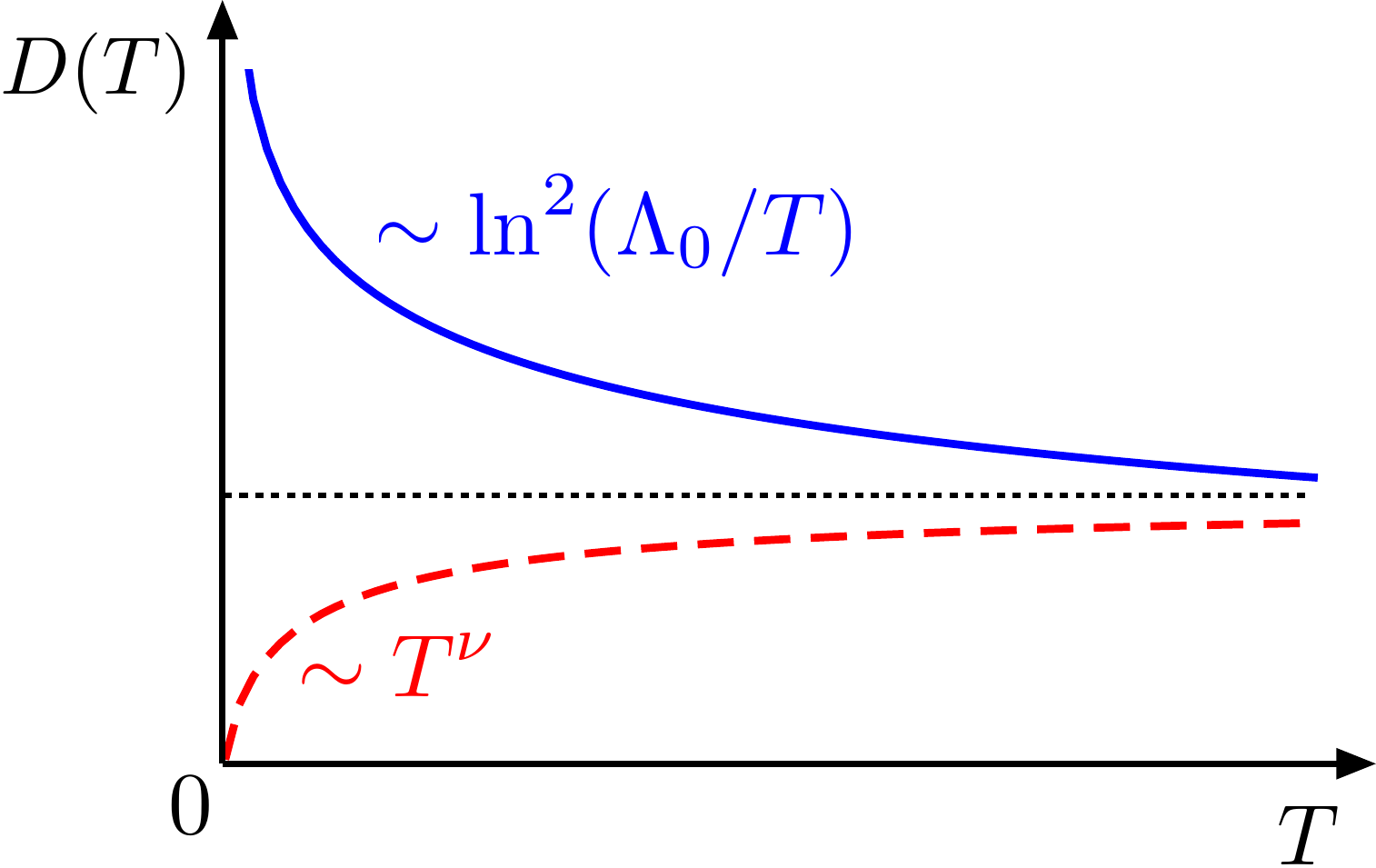}
\end{center}
\caption{(Color online) Sketch of the temperature dependence of the diffusion coefficient   for  $m_+>m_-$ (blue solid line) and $m_+<m_-$ (red dashed line). The dotted line indicates the scaling result $D(T)=\text{const.}$ expected at higher $T$. 
}\label{fig:diffusion}
\end{figure}

{\it Diffusion coefficient.}---We now turn to transport, which can be induced and probed in cold atoms by controlling the trapping potential or artificial gauge fields \cite{OttPRL2004,PalzerPRL2009,SchneiderNP2012,RonzheimerPRL2013,GoldmanRPP2014,ChienNP2015}. 
A particularly relevant response function is the  temperature-dependent diffusion coefficient $D(T) $ \cite{weiss1999quantum}. Experimentally, one can measure the diffusion coefficient  by preparing an initial state in which the impurity is   confined  by a trapping potential and watching the spreading of the wave packet after the trap is switched off \cite{CataniPRA2012,RonzheimerPRL2013,Kindermannarxic2016}. After sufficiently  long times, we expect diffusive   expansion  limited by inelastic collisions with the Fermi gas. The variance of the impurity position must then increase with time as $\sigma^2(t)=4D(T)t$. The diffusion coefficient  is connected with the impurity mobility $\mu(T)$   via Einstein's relation $D(T)=k_BT \mu(T)$ \cite{weiss1999quantum}.

For $m_+>m_-$, we   calculate  $D(T)$   in  the linear response regime as $D(T)=k_BT \tau_{\text{tr}}/M_0$, where $M_0=M(\Lambda_0)$ is the bare mass and $\tau_{\text{tr}}$ is the transport relaxation time of the long-lived quasiparticles. The transport relaxation time can be determined by a partial summation of diagrams in the Kubo formula for the mobility  \cite{LangrethPR1964,MahanPR166} or more simply using a memory function approach  \cite{GotzePRB1972} to lowest order in $\gamma$. We find \be
D(T)\propto1/\gamma^2(T)\propto \ln^2(\Lambda_0/T) \label{mobility1}.
\ee 
Note that from scaling  we would expect $D(T)=\text{const.}$ since the diffusion is limited by a marginal interaction. The enhancement of $D(T)$ by  the logarithmic correction in Eq. (\ref{mobility1}) is due to the flow of effective coupling constant to weak coupling as $T\to0$. 

For $m_+<m_-$, we can still use the memory function approach, which does not rely on the existence of quasiparticles. First, we employ the Lee-Low-Pines transformation \cite{LeePR1953,Grusdtarxiv20152} to the frame comoving with the impurity and obtain the transformed Hamiltonian \be
\tilde H_{\text{mob}}=\frac1{2M}\left(\mathbf P+i\int d^2r\, \Psi^\dagger \nabla \Psi\right)^2+H_{\text{loc}},\label{effectiveH}
\ee
where $\mathbf P$ is the total momentum of the system. The transformed impurity current operator is $\tilde {\mathbf J}_d =\frac1M\left(\mathbf P+i\int d^2r\, \Psi^\dagger \nabla \Psi\right)$. 
We  also  need the time derivative $\partial_t\tilde {\mathbf J}_d= -\frac{2\pi\gamma^*}{M}[\Psi^\dagger(0)\nabla\Psi(0)+\text{h.c.}]$.
The  mobility is   calculated using  $[\mu(T)]^{-1}= \frac{M^2}{\omega}\text{Im}[R_{\text{ret}}(\omega)-R_{\text{ret}}(0)]$. Here $R_{\text{ret}}(\omega)$ is the retarded memory function obtained  from the imaginary-time correlation $R(\tau)=\langle\partial_\tau\tilde {\mathbf J}_d(\tau)\cdot \partial_\tau\tilde {\mathbf J}_d(0)\rangle $  by Fourier transform and an analytic continuation to real frequency. Importantly, the impurity mobility is determined by a local correlation function  of the fermion field  which   can  be calculated at the low-energy fixed point described by Eq. (\ref{1Dchannels}) (see Supplemental Material \cite{supplem}).  We   find that for $m_+<m_-$ and $T\to 0$\be
D(T)\propto T^{\nu},\qquad \nu =1-\left(1-\frac{g}2\right)^2.\label{mobility2}
\ee
The  anomalous dimension  $\nu \sim \mc O(g)$ is analogous to the exponent of the Fermi edge singularity in  metals \cite{gogolin2004bosonization}. In  sharp contrast with Eq. (\ref{mobility1}), in this case $D(T)$ vanishes in the limit $T\to0$.  The result   is depicted in Fig. \ref{fig:diffusion}.  

We would like to emphasize that our work differs from previous  studies of  dissipative quantum dynamics where an impurity localization transition can be driven by varying the strength of the coupling to an Ohmic bath \cite{FisherZwergerPRB1985,weiss1999quantum}. In that case, the impurity-bath coupling is strictly relevant or irrelevant. In our case, both regimes can be realized at weak coupling and are   due to the effects of a marginal interaction. Note in particular that for $m_+<m_-$ the mobility \emph{diverges} in the low-temperature limit as $\mu(T)\propto T^{-1+\nu}$, with  $\nu$  governed by a marginal boundary operator.

{\it Conclusion.}---We have shown that mobile impurities coupled to a 2D Fermi gas with quadratic band touching exhibit  anomalous single-particle and transport properties. Controlling the particle-hole asymmetry of the fermionic bands allows one to tune between a regime of well-defined quasiparticles and another regime where the quasiparticle weight vanishes due to an emergent orthogonality catastrophe. Experiments that probe the   expansion dynamics of an  impurity wave packet  should observe a striking difference in the temperature dependence of the diffusion coefficient in these two regimes.

\acknowledgements 
We are grateful to F.  Brito for helpful discussions. This work is supported by CNPq (M.A.C., R.G.P.) and FAPESP (M.A.C.).

\bibliographystyle{aipnum4-1}

\bibliography{article1}

\appendix
\section{Supplemental Material}
\subsection{1. Renormalization group equations}

To derive the perturbative RG equations, we   use the  impurity Green's function  \be
G_d(k)=G_d(\mathbf k,i\omega)=\frac{Z}{i\omega-E(\mathbf k)}.
\ee
For the Fermi gas we define a matrix Green's function
\be
\mathbbm G=\left(\begin{array}{cc}
\mc G_{11}&\mc G_{12}\\
\mc G_{21}&\mc G_{22}
\end{array}\right),
\ee 
with components
\be
\mc G_{mn}(\mathbf p,\tau)=-\langle T_\tau \Psi^{\phantom\dagger}_m(\mathbf p,\tau) \Psi^{\dagger}_n(\mathbf p,0)\rangle,
\ee
where $m,n\in \{1,2\}$ are sublattice indices with $\Psi_{1}(\mathbf p)=a_\mathbf p$, $\Psi_{2}(\mathbf p)=b_\mathbf p$. The orthogonal transformation that diagonalizes $H_0(\mathbf p)$ in the continuum limit  is $\Psi_{m}(\mathbf p)=\sum_{\lambda}U_{m\lambda}(\mathbf p)  \Psi_{\lambda}(\mathbf p)$,  
where $\lambda\in \{+,-\}$ is a band index, and  \be
U(\mathbf p)=U(\theta_{\mathbf p})=\left(\begin{array}{cc}
\sin\theta_{\mathbf p}&\cos\theta_{\mathbf p}\\
-\cos\theta_{\mathbf p}&\sin\theta_{\mathbf p}
\end{array}\right).\label{Umatrix}
\ee

The RG   equation for the effective coupling constant $\gamma(\Lambda)$ is obtained by considering the four-point function\bea
\hspace{-0.2cm}\Pi(\{x_i\})&=&-\sum_n\langle T_\tau\Psi^{\phantom\dagger}_{n}(x_1) \Psi^{\dagger}_{n}(x_2)d(x_3)d^\dagger(x_4)\rangle,
\eea
where $x_i=(\mathbf r_i,\tau_i)$, and its Fourier transform\bea
\hspace{-0.2cm}\Pi(\{  p_i \})&=&\int \prod_{i=1}^4d^2r_id\tau_ie^{i\omega_i\tau_i }e^{-i\mathbf p_i\cdot \mathbf r_i}\Pi(\{x_i\}).
\eea
The result is of the form\bea
\Pi(\{  p_i \})&=&2\pi \Gamma(\{p_i\})  \text{Tr}\{\mathbbm G(p_1)\mathbbm G(p_2)\}G_d(p_3)G_d(p_4)\nonumber\\
&&\times \delta_{\sum_i\mathbf p_i,0} 2\pi \delta\left(\sum_i\omega_i\right).
\eea
where $\Gamma(\{p_i\}) $ is the effective vertex.  At tree level, we have $\Gamma(\{p_i\}) =\gamma$.
 The quantum correction at one-loop level is calculated using a  cutoff scheme that integrates out large-momentum states in the shell $K^\prime<|\mathbf k|<K$, with $K^\prime=K e^{-d\ell/2}$. Here we define $d\ell =d\Lambda/\Lambda=2dK/K$, where $\Lambda=K^2/2M_0$ is the energy cutoff that corresponds to the momentum cutoff $K$. The   particle and hole diagrams   in Fig. 2(a) of the main text yield  \bea
\Gamma^{(2)}_{\text{part}}(\{p_i\to0\})&=&-\frac{\mu_+\gamma^2 Z d\ell}4,\label{Pi2part}\nonumber\\  
\Gamma^{(2)}_{\text{hole}}(\{p_i\to0\})&=&\frac{  \mu_-\gamma^2Zd\ell}4.\label{Pi2hole}
\eea
This leads to the RG equation for $\gamma(\Lambda)$ in Eq. (4a).

\begin{figure}
  \begin{center}
\includegraphics*[width=.75\columnwidth]{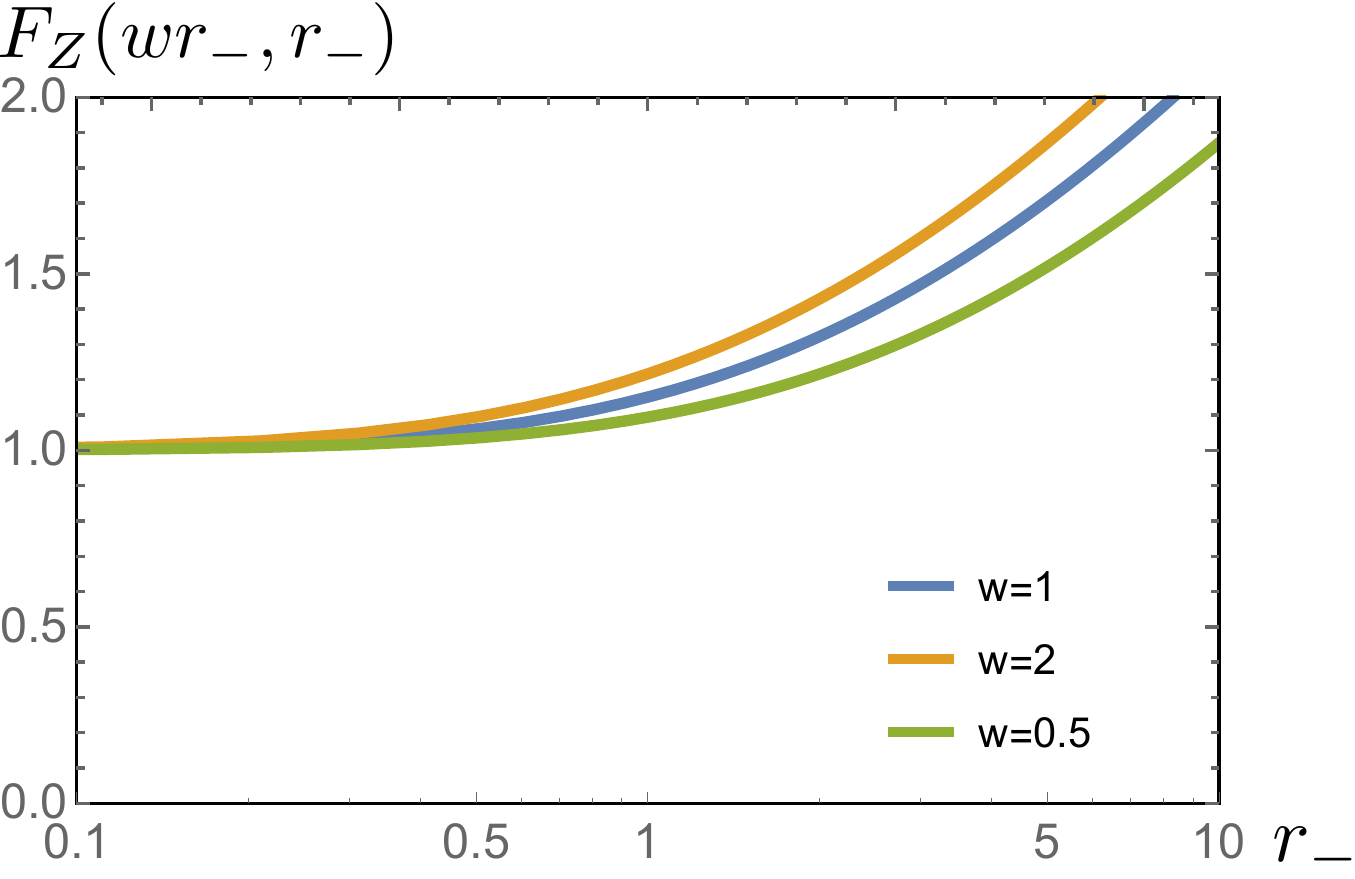}\\
\vspace{0.4cm}
\includegraphics*[width=.75\columnwidth]{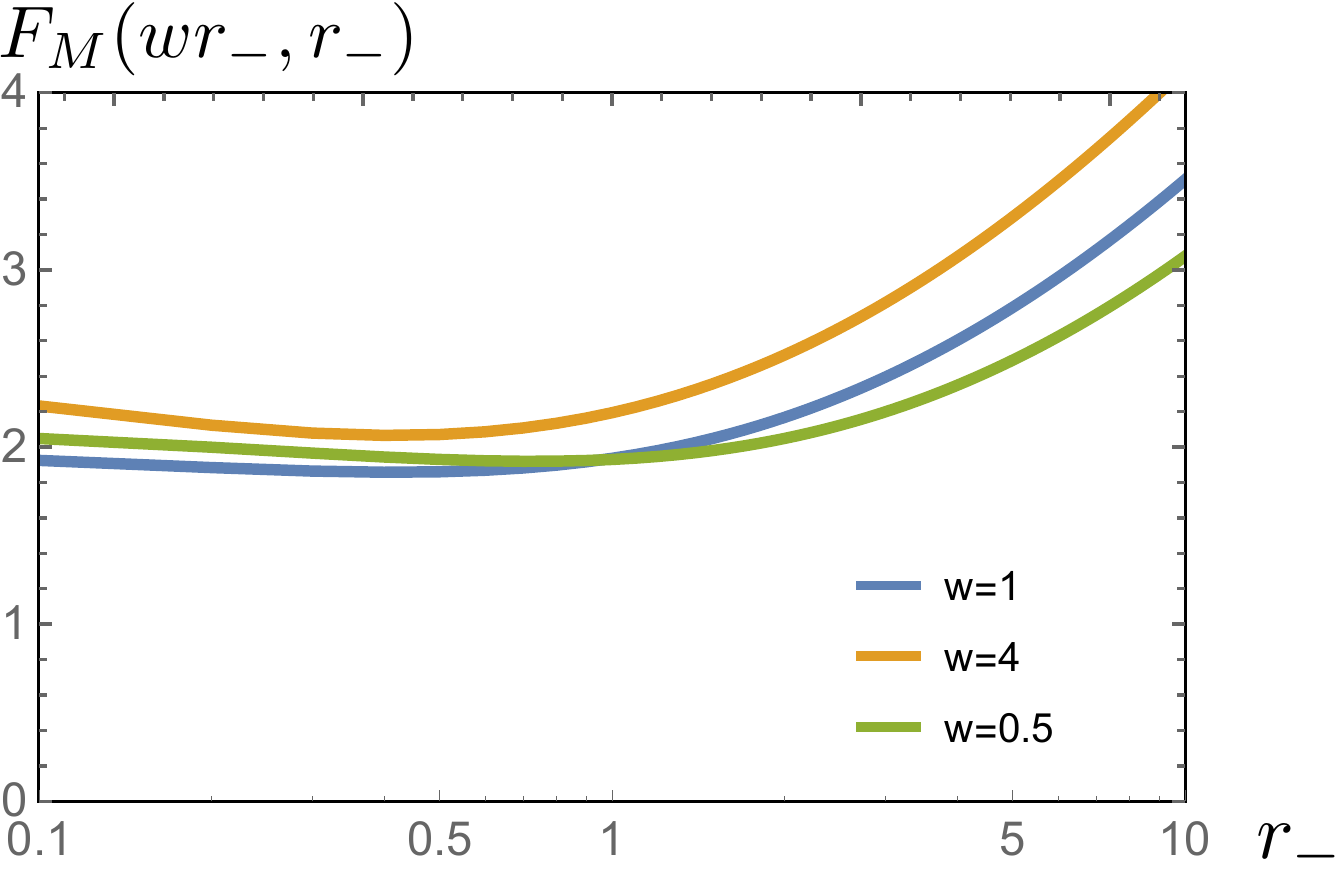}
\end{center}
\caption{
The functions of mass ratios   that appear in the RG equations for $M(\Lambda)$ and $Z(\Lambda)$. 
}\label{fig:functionF}
\end{figure}

To find the renormalization of the effective mass and quasiparticle weight, we calculate the two-loop ``sunrise'' diagram in the  impurity self-energy at order $\gamma^2$, given by
\bea
\Sigma^{(2)}(\mathbf p,i\omega)&=&-(2\pi \gamma)^2\sum_{\mathbf k}\sum_{\mathbf q}\int \frac{d\nu_{\mathbf k}}{2\pi}\int \frac{d\nu_{\mathbf q}}{2\pi}\nonumber\\
&&\times \text{Tr}\{\mathbbm G(k+q)\mathbbm G(k)\}G_d(p-q).
\eea
The integration   is performed such that both particle and hole momenta lie in the shell between $K^\prime$ and $K$.
We are only interested  in the result for   small $\omega, p$. To lowest order in $p$ and $\omega$   the self-energy has the form\bea
\Sigma^{(2)}(\mathbf p,i\omega)&\approx  & ia\omega -(a+b)\frac{p^2}{2M}.
\eea
We seek expressions for the coefficients $a$ and $b$, which determine the renormalized $M$ and $Z$ through $dZ=a$ and $dM=bM$. The result is 
\bea
a&=&-\frac{\gamma^2\mu_+\mu_-}2F_Z\left(\frac{m_+}{M},\frac{m_-}{M}\right)d\ell,\\
b&=&\frac{\gamma^2(\mu_+\mu_-)^{3/2}}{2M}F_M\left(\frac{m_+}{M},\frac{m_-}{M}\right)d\ell. 
\eea
The functions of mass ratios are  \begin{widetext}
\bea
F_Z\left(r_+,r_-\right)&=&\int_0^{\pi/2}d\alpha\, \frac{(1+r_+)(1+r_-)}{r_+r_-\sin\alpha\cos\alpha}\left[\frac{\frac{1+r_-}{2r_-}\cos^2\alpha+\frac{1+r_+}{2r_+}\sin^2\alpha}{\sqrt{\left(\frac{1+r_-}{2r_-}\cos^2\alpha+\frac{1+r_+}{2r_+}\sin^2\alpha\right)^2-  \sin^2\alpha\cos^2\alpha}}-1\right],\\
F_M(r_+,r_-)&=&2\int_0^{\pi/2}d\alpha\,\left[\frac{(1+r_+)(1+r_-)}{r_+r_-}\right]^{3/2}\frac{1-\sin(2\alpha)[3L+2(L^2-1)^{3/2}-2L^3]}{\sin^2(2\alpha)(L^2-1)^{3/2}},
\eea
\end{widetext}
where\be
L (\alpha,r_+,r_-)=\frac{\frac{1+r_+}{r_+}\sin^2\alpha+ \frac{1+r_-}{r_-}\cos^2\alpha}{\sin(2\alpha)}.
\ee
The functions $F_Z(r_+,r_-)$ and $F_M(r_+,r_-)$ are illustrated in Fig. \ref{fig:functionF}. Note that they are smooth functions of the mass ratios with values of order 1.

\section{2. Mapping onto 1D chiral fermions}

Using the partial wave expansion in Eq. (6) of the main text, we can write the free fermion model in the form\bea
H_0&=&\sum_{\lambda=\pm}\int\frac{d^2p}{(2\pi)^2}\, \frac{\lambda p^2}{2m_\lambda}\Psi^\dagger_\lambda(\mathbf p)\Psi^{\phantom\dagger}_\lambda(\mathbf p)\nonumber\\
&=&\sum_{\lambda,\ell}\frac{\lambda}{8\pi^2m_\lambda}\int_0^\infty dp \,  p^2   \chi^\dagger_{\lambda,\ell}(p)\chi^{\phantom\dagger}_{\lambda,\ell}(p).
\eea  
Performing the change of variable to $\mc K=p^2/(8\pi ^2 m_\lambda)$ and switching to the rescaled fields $\psi_{\lambda,\ell}(\mc K)$, we obtain \bea
H_0&=&\sum_{\lambda,\ell} \lambda \int_0^\infty d\mc K \, \mc K\psi^{\dagger}_{\lambda,\ell}(\mc K)\psi^{\phantom\dagger}_{\lambda,\ell}(\mc K)\nonumber\\
&=&\sum_\ell \int_{-\infty}^\infty d\mc K \, \mc K\psi^{\dagger}_{\ell}(\mc K)\psi^{\phantom\dagger}_{\ell}(\mc K),
\eea
where $\psi_{\ell}(\mc K)\equiv\theta(\mc K)\psi_{+,\ell}(\mc K)+i\text{sgn}(\ell)\theta(-\mc K)\psi_{-,\ell}(-\mc K)$. The chiral fermion in ``real'' space is the Fourier transform \be
\psi_\ell(x)=\frac1{\sqrt{2\pi}}\int_{-\infty}^{\infty}d\mc K\, e^{i\mc Kx}\psi_\ell(\mc K).
\ee

The potential scattering associated with the impurity at the origin can be written as\be
H_{\text{int}} =2\pi \gamma^*\sum_{\lambda,\mu}\int \frac{d^2k}{(2\pi)^2}\int \frac{d^2p}{(2\pi)^2}V^{\lambda,\mu}_{\mathbf k,\mathbf p}\Psi^\dagger_\lambda(\mathbf k)\Psi^{\phantom\dagger}_{\mu}(\mathbf p),\label{Hlocsup}
\ee
where $V^{\lambda,\mu}_{\mathbf k,\mathbf p}=[U^\dagger(\mathbf k)U(\mathbf p)]_{\lambda\mu}$. From Eq. (\ref{Umatrix}), we see that the amplitude $V^{\lambda,\mu}_{\mathbf k,\mathbf p}$ only depends on the polar angles of $\mathbf k$ and $\mathbf p$. Written as a matrix in the band basis:\be
V_{\mathbf k,\mathbf p}=\left(\begin{array}{cc}
\cos(\theta_{\mathbf k}-\theta_{\mathbf p})&\sin(\theta_{\mathbf k}-\theta_{\mathbf p})\\
-\sin(\theta_{\mathbf k}-\theta_{\mathbf p})&\cos(\theta_{\mathbf k}-\theta_{\mathbf p})
\end{array}\right).
\ee 
Substituting the partial wave expansion into Eq. (\ref{Hlocsup}), we obtain
\bea
H_{\text{int}} &=&\frac{\gamma^*}{2\pi}\sum_{\lambda,\mu}\int_0^\infty dk\,k\int_0^{\infty}dp\,p \int_{-\pi}^{\pi}\frac{d\theta_{\mathbf p}}{2\pi} \int_{-\pi}^{\pi}\frac{d\theta_{\mathbf k}}{2\pi}\nonumber\\
&&\times V^{\lambda,\mu}_{\mathbf k,\mathbf p} \sum_{\ell,\ell^\prime}\frac{e^{i(\ell \theta_{\mathbf p}-\ell^\prime \theta_{\mathbf k})}}{2\pi\sqrt{pk}}\,\chi^\dagger_{\lambda,\ell}(k)\chi^{\phantom\dagger}_{\mu,\ell^\prime}(p)\nonumber\\
&=&\frac{g}{2} \int_0^\infty d\mc K\int_0^{\infty}d\mc K^\prime\, \left[w^{\frac12}\psi^\dagger_{1}(\mc K)\psi^{\phantom\dagger}_{1}(\mc K^\prime)\right.\nonumber\\
&&+w^{-\frac12}\psi^\dagger_{1}(-\mc K)\psi^{\phantom\dagger}_{1}(-\mc K^\prime)+\psi^\dagger_{1}(\mc K)\psi^{\phantom\dagger}_{1}(-\mc K^\prime)\nonumber\\
&&\left.+\psi^\dagger_{1}(-\mc K)\psi^{\phantom\dagger}_{1}(\mc K^\prime)+(1\to-1)\right],
\eea
where $w=m_+/m_-$. We separate two terms  $H_{\text{int}}=H_{\text{int}}^{(s)}+H_{\text{int}}^{(a)}$ with\bea
H_{\text{int}}^{(s)}&=&\frac{g}{2} \int_{-\infty}^\infty d\mc K\int_{-\infty}^{\infty}d\mc K^\prime\, \left[ \psi^\dagger_{1}(\mc K)\psi^{\phantom\dagger}_{1}(\mc K^\prime)\right.\nonumber\\
&&\left.+(1\to-1)\right],\\
H_{\text{int}}^{(a)}&=&  \int_0^\infty d\mc K\int_0^{\infty}d\mc K^\prime\, \left[\kappa_+\psi^\dagger_{1}(\mc K)\psi^{\phantom\dagger}_{1}(\mc K^\prime)\right.\nonumber\\
&&\left.+\kappa_-\psi^\dagger_{1}(-\mc K)\psi^{\phantom\dagger}_{1}(-\mc K^\prime)+(1\to-1)\right],\label{Hintasym}
\eea
where $\kappa_+\equiv(w^{\frac12}-1)g/2$ and $\kappa_-\equiv(w^{-\frac12}-1)g/2$.
The term $H_{\text{int}}^{(s)}$ is the usual potential scattering and is local in the real 1D  space\be
H_{\text{int}}^{(s)}= \pi g[\psi^\dagger_1(0)\psi^{\phantom\dagger}_1(0)+\psi^\dagger_{-1}(0)\psi^{\phantom\dagger}_{-1}(0)].
\ee
The term $H_{\text{int}}^{(a)}$ is  due to the asymmetry between  particle-to-particle versus hole-to-hole scattering processes. We can show that this  term is irrelevant in the RG sense.
In terms of the chiral fermions, we  write 
\bea 
H_{\text{int}}^{(a)}&=&\frac{\kappa_+}{2\pi}\int dx_1  dx_2\,\frac{\psi_1^\dagger(x_1)\psi^{\phantom\dagger}_1(x_2)}{(x_1+i\eta)(x_2-i\eta)}\nonumber\\
&&+\frac{\kappa_-}{2\pi}\int dx_1  dx_2\,\frac{\psi_1^\dagger(x_1)\psi^{\phantom\dagger}_1(x_2)}{(x_1-i\eta)(x_2+i\eta)}\nonumber\\
&&+(1\to-1).\label{deltaHas}
\eea
We   employ Eq. (\ref{deltaHas}) to work out the renormalization  using the operator product expansion of the fermion fields in real space. The imaginary parts with $\eta\to0^+$ control the positions of poles in the integrations over  Euclidean spacetime. This way we obtain the RG equations to second order in $\kappa_\pm$ \be
\frac{d\kappa_+}{d\ell}=\kappa_+^2,\qquad \frac{d\kappa_-}{d\ell}=-\kappa_-^2.
\ee
Note that  $\kappa_+<0$ and $\kappa_->0$ since $w<1$ in the   regime $m_+<m_-$.  Thus, the particle-hole asymmetry in the potential scattering is irrelevant. However,  like other irrelevant perturbations, $H_{\text{int}}^{(a)}$ may renormalize the  scattering amplitude $g$ at the low-energy fixed point.  This effect   can be accounted for if we replace the  weak coupling expression $g=\gamma^*\sqrt{m_+m_-}$ (valid to first order in $\gamma^*$) by the relation $g=\delta/\pi$, where $\delta$ is    the exact  phase shift of the potential, as in the usual OC \cite{gogolin2004bosonization}.   Dropping $H_{\text{int}}^{(a)}$, we  obtain the    Hamiltonian in Eq. (7) of the main text.

\section{3. Correlations in the regime $m_+<m_-$} 
We bosonize the chiral fermions in the form\be
 \psi^{\phantom\dagger}_\ell(x)\sim e^{-i\sqrt{2\pi}\varphi_\ell(x)},
\ee
where $\varphi_\ell(x)$ is a chiral bosonic field that obeys $[\varphi_\ell(x),\partial_r\varphi_{\ell^\prime}(x^\prime)]=-i\delta_{\ell,\ell^\prime}\delta(x-x^\prime)$. The bosonized version of the Hamiltonian in Eq.  (7) of the main text  reads\bea
H_{\text{loc}}&=&\sum_{\ell\in\mathbb Z}\frac12\int_{-\infty}^{+\infty}dx\, (\partial_x\varphi_\ell)^2\nonumber\\
&&-g\sqrt{\frac{\pi}{2}}[\partial_x\varphi_1(0)+\partial_x\varphi_{-1}(0)].
\eea
We can eliminate the boundary operator by performing a unitary transformation \be
U=\exp\left\{-ig\sqrt{\frac{\pi}{2}}[\varphi_1(0)+\varphi_{-1}(0)]\right\},\label{unitaryshift}
\ee
which acts as\be
\tilde H_{\text{loc}}=U^\dagger H_{\text{loc}}U=\sum_\ell\frac12\int_{-\infty}^{+\infty}dx\, (\partial_x\varphi_\ell)^2=H_0.
\ee
The decay of the impurity Green's function is due to the OC when we change the boundary condition in the $\ell=\pm1$ channels. Let us denote the ground state of $H$ (in the presence of  the impurity) as $|\tilde 0\rangle=U|0\rangle$, where $|0\rangle$ is the ground state of $H_0$. Following Ref. \onlinecite{AffleckJPA1994}, we write the impurity Green's function as\bea
\langle d(t)d^\dagger(0)\rangle&=&\langle 0|U(t)U^\dagger(0)|0\rangle.
\eea
This is a correlation function of the vertex operator in Eq. (\ref{unitaryshift}) evolved with the noninteracting Hamiltonian $H_0$. Normal ordering the exponentials and using the Baker-Hausdorff formula $e^Ae^B=e^Be^Ae^{[A,B]}$ we obtain\be
\langle d(t)d^\dagger(0)\rangle\propto t^{-g^2/2}.
\ee

To calculate the mobility, first we have to account for the finite mass of the impurity before we set up the calculation of the correlation function at the low-energy fixed point. The physical picture here is that  when  the impurity is very heavy    the particle-hole excitations in the Fermi gas can follow its motion  adiabatically. Thus, it is convenient to transform to the impurity frame   and then determine the current decay through  the correlation of the cloud that dresses the impurity. 

We rewrite the time derivative of the current operator in the impurity frame in terms of chiral fermions:\bea
\partial_t\tilde {\mathbf J}_d&=&-\frac{2\pi\gamma^*}{M}[\Psi^\dagger(0)\nabla\Psi(0)+\text{h.c.}]\nonumber\\
&=&-\frac{i\gamma^*}{2\pi M}\sum_{\lambda}\int d^2k \int d^2p\,\mathbf p  \Psi^{\dagger}_\lambda(\mathbf k)\Psi^{\phantom\dagger}_\lambda(\mathbf p)+\text{h.c.}\nonumber\\
&=&-\frac{i8\pi^4 \gamma^*}{M}\sum_{\lambda}(2m^3_\lambda)^{1/2}\int_0^\infty d\mc K \int_0^\infty d\mc K^\prime\, \nonumber\\
&&\times\sqrt{\mc K^\prime}\psi^\dagger_{\lambda,0}(\mc K)\psi^{\phantom\dagger}_{\lambda,1}(\mc K^\prime)+(1\to -1)+\text{h.c.}.
\eea
Note that the operator now transfer fermions between $\ell=0$ and $\ell=\pm1$ channels. 

Let us consider the correlation in imaginary time \be
R(\tau)=\langle \partial_t\tilde{\mathbf J}_{d}(\tau)\cdot \partial_t\tilde{\mathbf J}_{d}(0)\rangle.\label{forcecorr}
\ee
We can calculate the correlation at zero temperature and later obtain the finite-temperature result using a conformal mapping. At zero temperature, $R(\tau)$ decays  as a power law with an exponent determined by the  scaling dimension of the operator $\partial_t\tilde{\mathbf J}_{d}$. Since we are only interested in the exponent and not on the prefactor, let us focus on the contribution from $\ell=1$\bea
R(\tau)&\sim&R_0(\tau)R_1(\tau),
\eea
where we used the decoupling of $\ell=0$ and $\ell=1$ channels to factorize the correlation into\bea
R_0(\tau)&=&\int_0^\infty d\mc K_1 \int_0^{\infty}d\mc K_2\,\langle \psi^{\phantom\dagger}_{0}(\mc K_1,\tau) \psi^{\dagger}_{0}(\mc K_2,0)\rangle,\\
R_1(\tau)&=& \int_0^\infty d\mc K_1 \int_0^{\infty}d\mc K_2\,\sqrt{\mc K_1\mc K_2}\nonumber\\
&&\times\langle\psi^{ \dagger}_{1}(\mc K_1,\tau) \psi^{\phantom\dagger}_{1}(\mc K_2,0)\rangle.
\eea
The $\ell=0$ channel is free and $\psi_0(x)$ has scaling dimension $1/2$. Thus, it is easy to check that $R_0(\tau)\sim 1/\tau$. For the $\ell=1$ channel, the correlation is affected by the boundary operator. The dominant long-time decay stems from the vicinity of $x=0$ and can be written in  the form\be
R_1(\tau)\sim  \frac1\tau\langle\psi^{\dagger}_{1}(0,\tau)\psi^{\phantom\dagger}_{1}(0,0)\rangle.
\ee
This correlation for the fermion field is analogous to that in   the optical absorption rate in the x-ray edge problem \cite{gogolin2004bosonization}. Unlike the impurity Green's function, in this case the correlation decays more slowly when we turn on the potential with $g>0$. We have\bea
\hspace{-0.5cm}\langle\psi^{ \dagger}_{1}(0,\tau) \psi^{\phantom\dagger}_{1}(0,0)\rangle&=&\langle0|  e^{H_0\tau}\tilde \psi^{\dagger}_{1}(0) e^{-H_0\tau}\tilde\psi^{\phantom\dagger}_{1}(0)  | 0\rangle,
\eea 
where the transformed field\be
\tilde \psi_1(0)\sim e^{-i\sqrt{2\pi}(1-g/2)\varphi_1(0)}
\ee
has scaling dimension $(1-g/2)^2/2$.
As a result, the correlation decays as \be
\langle\psi^{ \dagger}_{1}(0,\tau) \psi^{\phantom\dagger}_{1}(0,0)\rangle\sim \tau^{-1+\nu},\label{powerlaw}
\ee
with $\nu=1-(1-g/2)^2=g-g^2/4$.
 
Putting everything together, the correlation in Eq. (\ref{forcecorr}) at zero temperature decays as \be
R(\tau)\sim \tau^{-3+\nu}.
\ee
Mapping to finite temperature and real time, we obtain  \bea
\text{Im}R_{\text{ret}}(\omega)&\sim& \int_0^{\infty}dt\, e^{i\omega t}\left[\frac{\pi T}{\sinh(\pi T t)}\right]^{3-\nu}\nonumber\\
&\sim & T^{2-\nu}B\left(\frac{3-\nu}2-\frac{i\omega}{2\pi T},\nu-2\right),
\eea
where $B(x,y)$ is Euler's beta function.  Since we are interested in the dc limit, we expand the beta function for small $\omega$ and  obtain\be
\text{Im}R_{\text{ret}}(\omega)\sim \omega T^{1-\nu}.
\ee
This leads to the result for the mobility $\mu(T)\propto T^{-1+\nu}$ and for  the diffusion coefficient $D(T)\propto T^\nu$.


\end{document}